\documentclass[aps,twocolumn]{revtex4-2}

\usepackage{amsmath}
\usepackage{amsfonts}
\usepackage{graphicx}
\usepackage{braket}
\usepackage{enumitem}

\newcommand{\tr}{\text{tr}}

\begin{document}
\title{Optimized Correlation Measures in Holography}

\author{Newton Cheng}
\email{newtoncheng@berkeley.edu}
\affiliation{Center for Theoretical Physics and Department of Physics, University of California, Berkeley, CA, 94720, USA}

\date{\today}
\begin{abstract}
We consider a class of correlation measures for quantum states called optimized correlation measures, defined as a minimization of a linear combination of von Neumann entropies over purifications of a given state. Examples include the entanglement of purification $E_P$ and squashed entanglement $E_{\text{sq}}$. We show that when evaluating such measures on ``nice" holographic states in the large-$N$ limit, the optimal purification has a semi-classical geometric dual. We then apply this result to confirm several holographic dual proposals, including the $n$-party squashed entanglement. Moreover, our result suggests two new techniques for determining holographic duals: holographic entropy inequalities and direct optimization of the dual geometry.
\end{abstract}
\maketitle
\section{Introduction}
One of the most important tools in AdS/CFT is the RT/HRT prescription \cite{Ryu:2006bv, Hubeny:2007xt} for computing the von Neumann entropy of a boundary region $A$ in the state $\rho_{AB}$:
\begin{equation}\label{rt}
	S_A(\rho_{AB}) = \frac{\mathcal{A}[\mathcal{M}]}{4G_N},
\end{equation}
where $\mathcal{A}[\mathcal{M}]$ is the area of a homologous extremal surface anchored to the boundary of $A$ in the dual spacetime.  (\ref{rt}) provides an elegant interpretation of the bulk-boundary correspondence: an information-theoretic quantity about the boundary state can be computed as the area of a ``dual'' geometric object in the bulk, which is often simpler to do than working directly in the boundary theory. If $\rho_{AB}$ is pure, then $S_A$ captures the entire entanglement structure of the state. However, the von Neumann entropy is not the only correlation measure for a generic quantum state; indeed, it is well-known that the von Neumann entropy is a poor measure of entanglement once one considers mixed or $n$-party states with $n > 2$. Despite this, with a few exceptions, correlation measures beyond entropy have received little attention in a gravitational context. We therefore aim to study holographic duals for measures better suited to studying the entanglement structure of holographic mixed states, which are ubiquitous when studying e.g. black holes.

We will consider a class of correlation measures called \textit{optimized correlation measures}, whose bipartite forms were extensively studied in \cite{Levin:2019ab}. Given an arbitrary $n$-party state $\rho_{A_1\ldots A_n}$, these measures are defined to be a minimization of a function $f^{\alpha}$:
\begin{equation}\label{opt}
E^{\alpha}(A_1:\ldots :A_n) = \inf_{\ket{\psi}_{A_1\ldots A_nA_1'\ldots A_n'}}f^{\alpha}(\ket{\psi}\bra{\psi}),
\end{equation}
where the minimization is taken over all purifications $\ket{\psi}_{A_1\ldots A_nA_1'\ldots A_n'}$ of $\rho_{A_1\ldots A_n}$, so that $\tr_{A_1'\ldots A_n'}\ket{\psi}\bra{\psi} = \rho_{A_1\ldots A_n}$. Denoting the set of all $2^{2n}-1$ non-empty combinations of these regions by $\mathcal{R}$, the function $f^{\alpha}$ is a linear combination of von Neumann entropies on the extended Hilbert space:
\begin{equation}
f^{\alpha}(\ket{\psi}\bra{\psi}) = \sum_{\mathcal{J}\in \mathcal{R}}\alpha_{\mathcal{J}}S_{\mathcal{J}}(\ket{\psi}\bra{\psi}),
\end{equation}
where $\alpha \in \mathbb{R}^{2^{2n}-1}$ and the sum runs over all elements $\mathcal{J}$ of $\mathcal{R}$. As in \cite{Levin:2019ab}, we will impose a non-negativity condition $\sum_{\mathcal{J}}\alpha_{ \mathcal{J}} \geq 0$, or else choosing part of the extension to be maximally mixed on an enormous Hilbert space gives $E^{\alpha} \to -\infty$. An example of such a measure that has recently garnered significant interest is the entanglement of purification \cite{doi:10.1063/1.1498001}:
\begin{equation}
E_P(A:B) = \min_{\ket{\psi}_{AA'BB'}}S_{AA'}.
\end{equation}
for which $\alpha_{AA'} = 1$ and all other $\alpha_{ \mathcal{J}} = 0$. For holographic states, $E_P$ is conjectured to be dual to a particular minimal surface in the bulk called the entanglement wedge cross-section $E_W$ \cite{Takayanagi:2017knl, Nguyen:2017yqw, Bao:2018gck, Umemoto:2018jpc}. The
$E_P = E_W$ conjecture, as well as related proposals for the dual to $E_W$, has since attracted significant attention \cite{Hirai:2018jwy,Espindola:2018ozt,Agon:2018lwq,Caputa:2018xuf,Ghodrati:2019hnn,BabaeiVelni:2019pkw,Jokela:2019ebz,Du:2019emy,Bao:2019wcf,Harper:2019lff,Dutta:2019gen,Jeong:2019xdr,Bao:2019zqc,Chu:2019etd,Tamaoka:2018ned,Kudler-Flam:2018qjo,Kudler-Flam:2019oru,Kudler-Flam:2019wtv,Kusuki:2019zsp,Kusuki:2019rbk,Kusuki:2019evw}. Other examples include the $R$ and $Q$-correlation introduced in \cite{Levin:2019ab} and the squashed entanglement $E_{\text{sq}}$ \cite{doi:10.1063/1.1643788}, which we discuss in the main body of the paper.

A generic correlation measure of the form  (\ref{opt}) is generally very difficult to compute, due to the enormous parameter space of purifications that that needs to be searched. As a result, finding the holographic duals to such measures can be similarly difficult, relying on indirect evidence such as consistency checks with inequalities, or requiring strong assumptions like the surface-state correspondence \cite{Miyaji:2015yva}. Holographic duals to other optimized correlation measures have been proposed \cite{Hayden:2011ag, Umemoto:2019jlz, Levin:2019krg}, but their conclusions rely on the assumption that the optimal purification in (\ref{opt}) can be taken to be geometric, in the sense that the purification lives in the boundary CFT Hilbert space and is dual to a semi-classical geometry. The question of the assumption's validity was first posed in \cite{Hayden:2011ag} in the context of the squashed entanglement $E_{\text{sq}}$, and has since remained an open problem in holography. Some recent progress towards an answer appeared in \cite{Bao:2019wcf}, where it was argued that geometric minimization was sufficient to compute the bipartite $E_P$, and hence one could employ the bit thread formulation of holography \cite{Freedman:2016zud} on the purifying state to provide evidence for $E_P = E_W$. 

In this work, we argue that, up to reasonable assumptions we will outline below, something even stronger holds: for any $n$-party optimized correlation measure evaluated on ``nice" holographic states, minimizing over the subset of geometric purifications is sufficient to achieve the minimum over all purifications. Using our result, we confirm the holographic proposals in \cite{Hayden:2011ag, Umemoto:2019jlz, Levin:2019krg}, and in particular, we find a dual to the $n$-party squashed entanglement. Moreover, our result suggests that holographic duals can be determined by directly optimizing the purified geometry, as opposed to relying on indirect evidence such as obeying the same set inequalities.

Throughout this paper, we will work at leading order in the large-$N$, large central charge limit. We will also assume that a suitable regularization scheme is in place, e.g. a radial cutoff, such that all the correlation measures of interest are finite and Hilbert spaces can be tensor-factorized. We will moreover assume that the purifying subsystem is finite dimensional and, crucially, that the infimum in  (\ref{opt}) can turned into a minimum; this is an open question for generic quantum states, but we argue below that this is not a major obstacle for holographic states.

The paper is organized as follows: in Section 2, we review the necessary assumptions and properties of holographic states to prove our claim. In Section 3, we apply the properties of holographic states to prove the sufficiency of minimizing over states with semi-classical duals. In Section 4, we discuss applications of the result to holographic systems and determining duals to various correlation masures. Finally, in Section 5, we discuss potential avenues of future work for understanding holographic states.

To make notation cleaner we will use $U = \bigcup_{i=1}^nA_i$ to refer to the union of all $n$ parties $A_1,\ldots, A_n$ of an $n$-party state, and use $U' = \bigcup_{i=1}^nA_i'$ to refer to the purifying subsystem, e.g. $\rho_{A_1\ldots A_n} = \rho_U$. We will also draw a distinction between \emph{entanglement measures} and \emph{total correlation measures}; entanglement measures are intended to capture purely quantum correlations, while the more general total correlation measures are allowed to capture a combination of quantum and classical correlations. Entanglement measures are usually required to satisfy more stringent properties that are not satisfied by a generic $E^{\alpha}$, as we discuss below. Finally, although we use the term ``minimal surface,'' our arguments do not rely on time symmetry, and we therefore expect no issues to arise in considering dynamical setups with extremal surfaces.

\section{Preliminaries}
\subsection{Correlation measures}
We first briefly introduce the general, axiomatic approach to correlation measures in quantum information. We are given a functional $E$ defined on the set of density matrices $\mathcal{D}(\mathcal{H})$ of a Hilbert space $\mathcal{H}$, and in order for $E$ to be a good correlation measure, we require that $E$ satisfy some given set of axioms that are believed to be reasonable. Arguably the most important axiom for an entanglement measure is monotonicity under local operations and classical communication (LOCC):
\begin{equation}
    E(\rho) \geq E(\Lambda(\rho)),
\end{equation}
for any LOCC operation $\Lambda$, as entanglement nevers increase under LOCC. Other axioms that are usually enforced for entanglement measures include that they should vanish on separable states, be continuous in the asymptotic regime corresponding to $n\to \infty$ copies of a given state, and be normalized in the sense that $E(\Phi_+^d) = \log d$ for the maximally mixed state $\Phi_+^d$ of size $d$. One can show that any such $E$ that satisfies these axioms and is monotonic under LOCC reduces to the von Neumann entropy on pure states, singling it out as the unique (bipartite) entanglement measure for pure states \cite{doi:10.1063/1.1495917}.

The converse is not true: a correlation measure $E$ that reduces to the von Neumann entropy on pure states is not generically a good entanglement measure. However, such measures are still of interest if they are monotonic under LO -- these are total correlation measures, which may be sensitive to both quantum and classical correlations in a given state, rather than just quantum correlations from entanglement. The dependence on classical correlations allows these $E$ to increase under CC. The most well-known example of such a measure is the bipartite mutual information:
\begin{equation}
	I(A:B) = S_A + S_B - S_{AB},
\end{equation}
which captures all correlations, quantum and classical, between $A$ and $B$ \cite{PhysRevA.72.032317}, and reduces to $S_A$ for a pure state. In other words, $S_A$ is really the unique correlation measure for bipartite pure states in general. $E_P$ is also an example of a total correlation measure, as it is monotonic under LO (but not CC) and trivially reduces to $S_A$ on pure states.

While the story is very clean for bipartite, pure states, the situation is unclear for any other kind of state. Consider, for instance, a generic mixed state $\rho_{AB}$. Then we will generally find:
\begin{equation}
	S_A(\rho_{AB}) \neq S_B(\rho_{AB}),\quad S_{AB}(\rho_{AB}) \neq 0
\end{equation}
so $S_A$ is not only capturing the entanglement between $A$ and $B$, but can be seen as receiving contributions from the entanglement between $A$ and a purifying subsystem. The search for good entanglement measures for mixed states remains an active subfield of quantum information; see e.g. \cite{RevModPhys.81.865} for a discussion on various, generally inequivalent, measures. The inequivalence corresponds to differing operational interpretations. Moreover, $S_A$ is a bipartite correlation measure, and cannot capture intrinsically $n$-party entanglement for $n > 2$. The simplest example is the GHZ$_3$ state:
\begin{equation}
    \ket{\text{GHZ}_3}_{ABC} = \frac{1}{\sqrt{2}}(\ket{000}+\ket{111}),
\end{equation}
with the party labels corresponding to each qubit. The reduced density matrix of any 2 qubits is
\begin{equation}
    \tr_A\ket{\text{GHZ}_3}\bra{\text{GHZ}_3}_{ABC} = \frac{1}{2}(\ket{00}\bra{00} + \ket{11}\bra{11}),
\end{equation}
which is a separable state, and hence contains no entanglement between qubits $BC$. The entropy $S_A = \log 2$ is only able to capture the bipartite correlations between $A$ and $BC$, rather than the intrinsic tripartite correlations.

In holography, we are interested in the theory of correlation measures because mixed states and multipartite states are ubiquitous, e.g. a black hole with inverse temperature $\beta$ or multiple boundary subregions, and the issues with the von Neumann entropy mean studying the entanglement structure of such states is difficult. However, just as the von Neumann entropy has a geometric dual, we expect that other information-theoretic quantities defined on the boundary state should have a geometric dual. Understanding the dual objects can be extremely useful, because there are situations where the geometric description is easier to work with than the CFT description. A simple example is computing the entropy of a single interval in the vacuum state of a holographic 2D CFT: the CFT calculation, while not particularly complicated, is more work than computing the area of a geodesic in AdS$_3$. Indeed, $E^{\alpha}$ is very difficult to compute for an arbitrary quantum state; $E_P$ is one of the simplest examples of an optimized correlation measure and already suffers from this issue. The geometric dual generally makes these measures easier to compute or understand, which can lead to insight into the fine-grained structure of holographic states. Conversely, holographic systems are a useful testing ground for quantum information, precisely because we are able to understand quantities that are usually extremely complicated for generic quantum states.

\subsection{Linearity and geometric purifications}
To make our argument, we will need to employ two crucial properties of holographic states. The first property we will use is that entropies in holography are linear at leading order in the $1/N$ expansion \cite{Almheiri:2016blp}. More precisely, given a superposition of holographic states $\rho_{AB} = \sum_i^Mp_i\rho^i_{AB}$ with exponentially-suppressed overlap (as is the case for states with macroscopically-distinct dual geometries), the entropy of the $A$ subregion is approximately
\begin{equation}\label{lin}
	S_A(\rho_{AB}) = \sum_i^Mp_iS_A(\rho^i_{AB}) + S_{\text{mix}} +\ldots,
\end{equation}
where $S_{\text{mix}} = -\sum_i^Mp_i\log p_i$ is the entropy of mixing, and the ellipses indicate terms of order $O(1/N^2)$ and smaller. The assumptions made in computing  (\ref{lin}) break down when $M \sim e^{O(c)}$, with the most apparent effect being that $S_{\text{mix}}$ can become leading order.

Now let $\ket{\psi}_{ABA'B'}$ be a purification of $\rho_{AB}$ defined on the entire boundary. The second property we will use is that we can expand $\ket{\psi}$ as a superposition of states $\{\ket{\phi_i}\}$ dual to semi-classical geometries:
\begin{equation}\label{pure}
	\ket{\psi} = \sum_i^M c_i\ket{\phi_i},
\end{equation}
such that $\{\ket{\phi_i}\}$ are themselves purifications of $\rho_{AB}$ \cite{Bao:2019wcf}:
\begin{equation}
	\tr_{A'B'}\ket{\phi_i}\bra{\phi_i}_{ABA'B'} = \rho_{AB},
\end{equation}
for all $\ket{\phi_i}$. Because our result relies so crucially on this result, we reproduce the argument from \cite{Bao:2019wcf} for the existence of such a decomposition here with added detail and attention paid to the necessary assumptions.

Given a holographic state $\rho_{AB}$, let $\ket{\psi}_{ABA'B'}$ be any purification on the boundary with a geometric dual. We are free to act on the purifying subsystem $\mathcal{H}_{A'B'}$ with any operation that preserves the purity of the overall state; this is the standard statement from quantum mechanics that all purifications of a given state in the same Hilbert space are related by a unitary acting on $\mathcal{H}_{A'B'}$. In particular, we are free to insert e.g. pairs of black holes using the gluing procedure of \cite{Engelhardt:2018kcs}. Let $\{\ket{\phi_i}\}$ be the set of purifications generated by this procedure -- these states all live in the holographic CFT Hilbert space and will generally be a proper subset of the set of all geometric purifications.

We now make a reasonable assumption: we can fill the purifying subsystem with black hole microstates. That is, because we are free to tune the parameters of the black holes in the purifying subregion to fill very small or very large regions, such states should have support over any complete basis of $\mathcal{H}_{A'B'}$. Therefore, given an arbitrary basis $\ket{\beta_i}$ of $\mathcal{H}_{A'B'}$, we should be able to write
\begin{equation}
    \sigma_i = \tr_{AB}\ket{\phi_i}\bra{\phi_i} =  \sum_{j}c_{ij}\ket{\beta_j}\bra{\beta_j} + \sum_{j}c_{ijk}\ket{\beta_j}\bra{\beta_k}.
\end{equation}
We now make another assumption: the black hole microstates we consider are typical in the sense that there is a basis in which the $c_{ijk}$ are exponentially suppressed in $N$ and $c_{ij}$ is approximately invertible; for example, we could consider the eigenstate thermalization hypothesis for the energy eigenbasis. For two $\sigma_i,\sigma_j$ with $i\neq j$, their bulk geometries in $\mathcal{H}_{A'B'}$ are macroscopically distinct, and hence $\sigma_i,\sigma_j$ are approximately orthogonal up to exponential corrections. Then any superposition of the $\ket{\phi_i}$ will also be a purification of $\rho_{AB}$. The invertibility of $c_{ij}$ implies we can also invert the $\sigma_i$, so that
\begin{equation}
    \ket{\beta_j}\bra{\beta_j} \sim \sum_i c_{ij}^{-1}\sigma_i,
\end{equation}
which implies that there are at least $\dim \mathcal{H}_{A'B'}$ many linearly independent $\sigma_i$. In other words, we are able to construct any density matrix in the purifying subsystem as a superposition of geometric states on the subsystem. Again invoking the unitary equivalence of purifications, we conclude that, up to exponential corrections, any arbitrary purification on the boundary can be expanded as a superposition of geometric purifications.

We will assume the requisite assumptions are satisfied to use the above properties: we do not work beyond leading order in $1/N$, we work with an appropriately dense and typical subspace of black hole microstates, and the number of terms $M$ that appear in  (\ref{pure}) is exponentially suppressed relative to the full Hilbert space dimension. In other words, the results of this paper will only apply to suitably ``nice" holographic states. Note, however, that these assumptions are common in much of the literature and much of the standard lore in holography similarly only holds for such ``nice" states; for example, it is known that the error correction properties of AdS/CFT, and hence other properties, can be radically changed with large $M$ e.g. \cite{Akers:2019wxj}. Indeed, we do not expect the results of this work to apply beyond these ``nice" states.

\section{Minimizations over geometric extensions}
We now present the main result of this paper: when evaluating a correlation measure $E^{\alpha}$ of the form  (\ref{opt}) on a holographic state, it is sufficient to minimize only over purifications that are geometric, so that the purifying subsystem corresponds to a spatial region on the boundary.

Suppose we are interested in computing some $n$-party optimized correlation measure $E^{\alpha}$ for an arbitrary $n$-party CFT state $\rho_U = \rho_{A_1\ldots A_n}$ on a Hilbert space $\mathcal{H} = \otimes\mathcal{H}_n$ that is a subsystem of the full boundary Hilbert space $\mathcal{H}_{\text{CFT}}$. Let $\ket{\psi}_{UU'}$ be a purification of $\rho_{U}$, and moreover, let us choose $\ket{\psi}_{UU'}$ such that it achieves the desired minimum over all purifications:
\begin{equation}
	E^{\alpha}(A_1:\ldots:A_n) = f^{\alpha}(\ket{\psi}\bra{\psi}) = \sum_{\mathcal{J}\in \mathcal{R}}\alpha_{ \mathcal{J}}S_{\mathcal{J}}(\ket{\psi}\bra{\psi}),
\end{equation}
For our arguments to hold, we will need to assume that $\ket{\psi}_{UU'}$ exists and lives in the large-$N$ CFT Hilbert space $\mathcal{H}_{\text{CFT}}$. This is not immediately obvious, given that the purifying subsystem may have unbounded, but finite, dimension. As a heuristic argument for why these assumptions are reasonable, we note that we are free to tune our regulator to make $\mathcal{H}$ have arbitrarily large, but finite, dimension. A caveat is that we are ultimately limited by a minimal cutoff scale that vanishes in the infinite-$N$ limit. Modulo this caveat, the idea is then to keep tuning $\mathcal{H}$ until it can fit such a $\ket{\psi}_{UU'}$. For a particular choice of $\alpha$, it may be possible that there exists a suitably finite bound on the dimension of the purifying subsystem, in which case this assumption is unnecessary. This is the case for $E_P$.

Now expand $\ket{\psi}_{UU'}$ in the basis of holographic states with semi-classical duals, with the added property that each state in the expansion also be a geometric purification of $\rho_U$:
\begin{equation}\label{expan}
	\ket{\psi}_{UU'} = \sum_i^Mc_i\ket{\phi_i}_{UU'},
\end{equation}
such that $\tr_{U'} \ket{\phi_i}\bra{\phi_i}_{UU'} = \rho_{U}$ for all $i$. Writing $p_i = |c_i|^2$, we can lower bound $E^{\alpha}$ as follows:
\begin{align}
	E^{\alpha}(A_1:\ldots:A_n) &= f^{\alpha}(\ket{\psi}\bra{\psi}) \\
	&= \sum_\mathcal{J\in\mathcal{R}}\alpha_{\mathcal{J}}S_{\mathcal{J}}(\ket{\psi}\bra{\psi}) \\
	&\sim \sum_\mathcal{J\in\mathcal{R}}\alpha_{\mathcal{J}}\sum_i \bigg[{p_i}S_{\mathcal{J}}(\ket{\phi_i}\bra{\phi_i}) + S_{\text{mix}}\bigg] \\
	&\geq \min\limits_i\sum_\mathcal{J\in\mathcal{R}}\alpha_{\mathcal{J}}S_{\mathcal{J}}(\ket{\phi_i}\bra{\phi_i}) .
\end{align}
Note that $\left(\sum_{\mathcal{J}\in\mathcal{R}}\alpha_{\mathcal{J}}\right)S_{\text{mix}} \geq 0$ as both terms are non-negative. By the minimality of our choice of $\ket{\psi}_{UU'}$, we must also have the upper bound
\begin{equation}
	\min\limits_i\sum_\mathcal{J\in\mathcal{R}}\alpha_{\mathcal{J}}S_{\mathcal{J}}(\ket{\phi_i}\bra{\phi_i})  \geq E^{\alpha}(A_1:\ldots:A_n).
\end{equation}
We therefore find
\begin{equation}
	E^{\alpha}(A_1:\ldots:A_n) = \min\limits_i\sum_\mathcal{J\in\mathcal{R}}\alpha_{\mathcal{J}}S_{\mathcal{J}}(\ket{\phi_i}\bra{\phi_i}) 
\end{equation}
for some geometric purification $\ket{\phi_i}_{UU'}$ of the original state $\rho_U$. We have hence found that the minimum over all purifications can always be achieved, to leading order, by only considering geometric purifications. 

Many correlation measures of interest are usually defined as optimizations over all \textit{extensions} of a given state, rather than purifications. An extension of a state $\rho_U$ is simply any state $\rho_{UE}$ in an extended Hilbert space $\mathcal{H}_U\otimes\mathcal{H}_{E}$ such that $\tr_E\rho_{UE} = \rho_U$. Such correlation measures have the form:
\begin{equation}
E^{\alpha}(A_1:\ldots :A_n) = \inf_{\rho_{UE}}\sum_{\mathcal{J}\in \mathcal{R}}\alpha_{ \mathcal{J}}S_{\mathcal{J}}(\rho_{UE}),
\end{equation}
where $\mathcal{R}$ is the order $2^{n-1}-1$ set of all non-trivial combinations of $A_1,\ldots, A_n, E$. However, it is clear that this is entirely equivalent to the definition in  (\ref{opt}), as purifying the $\rho_{UE}$ to $\ket{\psi}_{UEE'}$ leaves all the entropies appearing on the RHS unchanged, and hence the optimization will yield the same result. The difference lies only in the redundancy of the search. Using $E_P$ as an example, we could equivalently write its definition as:
\begin{align}
    E_P(A:B) &= \min_{\rho_{ABA'}}S_{AA'}(\rho_{ABA'}) \\
    &= \min_{\ket{\psi}_{ABA'B'}}S_{AA'}(\ket{\psi}\bra{\psi}_{ABA'B'}).
\end{align}
Therefore, our result also applies to correlation measures defined in this way. For the sake of clarity with respect to other literature, we will always use the most common definitions of correlation measures, even if such definitions are not written as optimizations over purifications.

\section{Applications in holography}
\subsection{Optimizing geometries}
Our result find immediate application in the search for holographic duals to $E^{\alpha}$, whereby we can directly optimize the appropriate combination of surfaces dual to the desired linear combination of entropies on a given geometry. Two recent examples are the $Q$-correlation $E_Q$ and $R$-correlation $E_R$ \cite{Levin:2019ab}, defined as:
\begin{align}
E_Q(A:B) &= \frac{1}{2}\min\limits_{\rho_{ABE}}\big[S_A + S_B + S_{AE} - S_{BE}\big] \\
&= \frac{1}{2}\min\limits_{\ket{\psi}_{ABA'B'}}\left[S_A + S_B + S_{AA'} - S_{BA'}\right], \\
E_R(A:B) &= \frac{1}{2}\min\limits_{\rho_{ABE}}\big[S_{AB} + 2S_{AE} - S_{ABE} - S_{E}\big] \\
&= \frac{1}{2}\min\limits_{\ket{\psi}_{ABA'B'}}\big[S_{AB} + 2S_{AA'} - S_{ABA'} - S_{A'}\big].
\end{align}
The interest in these measures stems form the fact that not all correlation measures of the form (\ref{opt}) are good correlation measures in the sense that $E^\alpha$ are generically not monotonic under LO. \cite{Levin:2019ab} found that enforcing the monotonicity condition, along with the non-negativity condition $\sum_{\mathcal{J}\in\mathcal{R}}\alpha_{\mathcal{J}} \geq 0$, leads to the intersection of convex cones in $\alpha$-space whose extreme rays include $E_P$, $E_Q$, $E_R$, and the squashed entanglement $E_{\text{sq}}$, which we discuss in the next subsection.

In \cite{Umemoto:2019jlz,Levin:2019krg}, holographic duals to $E_Q$ and $E_R$ were proposed, but with the necessary assumption that the minimization over geometric purifications is sufficient to achieve a minimum over all purifications. With this assumption, one finds that $E_Q$ is dual to the \emph{entanglement wedge mutual information} $E_M$ defined by
\begin{equation}
    E_M(A:B) = \frac{1}{2}\min\limits_{A'B'}\big[S_A + S_B + S_{AA'} - S_{BA'}\big],
\end{equation}
which is nothing more than the statement of $E_Q$, but with the minimization over geometric regions $A',B'$ that purify $\rho_{AB}$ and the entropies interpreted as minimal surfaces. Such an expression is useful precisely because, as described in \cite{Umemoto:2019jlz}, there are cases where the geometric description of the system is particularly simple, e.g. two disjoint regions in AdS$_3$, so that $E_M$ can be understood intuitively, and is even tractable to compute. Similarly, one finds the holographic dual to $E_R$:
\begin{align}
    E_R(A:B) &= \frac{1}{2}\min\limits_{A'B'}\big[S_{AB} + 2S_{AA'} - S_{ABA'} - S_{A'}\big] \\
    &= E_W(A:B),
\end{align}
which reduces to nothing more than the entanglement wedge cross-section, adding $E_R$ to a growing list of candidate information duals to $E_W$.

Our result confirms the results above, as we have shown that the crucial assumption regarding geometric minimizations is also valid. More generally, we can make a proposal for the holographic dual of any $E^\alpha$ as simply
\begin{equation}
    E^{\alpha}(A_1:\ldots:A_n) = \min_{A_1'\ldots A_n'}f^{\alpha}(\ket{\psi}\bra{\psi}),
\end{equation}
with the $A_1'\ldots A_n'$ interpreted as geometric regions that purify the original state. Again, the utility of such an expression is that there likely exist situations where the geometric description is relatively simple, making $E^{\alpha}$ easier to understand or compute.

We remark here that applying this result to the bipartite $E_P = E_W$ conjecture gives
\begin{equation}
    E_P(A:B) = \min_{A'B'}S(AA') = E_W(A:B),
\end{equation}
so that $E_W$ can be interpreted as a single von Neumann entropy in some purified geometry. This is very similar to an independent proposal for the information-dual of $E_W$ given in \cite{Dutta:2019gen}, where a simple von Neumann entropy, called the reflected entropy $S_R$, defined on a canonical purification of the given state is dual to $2E_W$. Understanding the relationship between the geometries dual to the purifications in $E_P$, $E_R$, and $S_R$ could provide insight into why multiple distinct correlation measures reproduce the same value in holography.

\subsection{Holographic entropy inequalties}
As another method of determining holographic duals, our result permits the use of holographic entropy inequalities in computing optimized correlation measures. One example of note is the squashed entanglement $E_{\text{sq}}$, defined by \cite{Tucci:2002ab,doi:10.1063/1.1643788}:
\begin{align}\label{sqentdef}
    E_{\text{sq}}(A:B) &= \frac{1}{2}\min\limits_{\rho_{ABE}}I(A:B|E) \\
    &= \frac{1}{2}\min\limits_{\rho_{ABE}}(S_{AE} + S_{BE} - S_{ABE} - S_{E})
\end{align}
The squashed entanglement has been extensively studied in the context of quantum information \cite{PhysRevA.69.022309,Brandao2011,Li2018,PhysRevLett.104.240405,Wilde2016,6832533,Pirandola:2017abc,Cope_2018}, as it satisfies several abstract properties that make it arguably the most promising measure of purely quantum correlations between parties in mixed states. For example: if $\rho_{AB}$ is a pure state, then the only extensions of the state are tensor products $\rho_{ABE} = \rho_{AB} \otimes \rho_E$, so
\begin{equation}
    \frac{1}{2}I(A:B|E) = \frac{1}{2}I(A:B) = S_A,
\end{equation}
and $E_\text{sq} = S_A$, reducing to the von Neumann entropy on pure states. Moreover, it was proven in \cite{Brandao2011} that $E_{\text{sq}}$ is faithful, i.e. $E_{\text{sq}}= 0$ if and only if $\rho_{AB}$ is a separable state. By taking $E = \emptyset$, one obtains the upper bound
\begin{equation}
    \frac{1}{2}I(A:B) \geq E_{\text{sq}}(A:B),
\end{equation}
which is easily interpreted: there cannot be more quantum correlations than total correlations. 

While $E_{\text{sq}}$ is difficult to compute for generic quantum states, it was observed in \cite{Hayden:2011ag} that the computation of $E_{\text{sq}}$ for holographic states could potentially be made simple because holographic entropy satisfies the monogamy of mutual information (MMI):
\begin{equation}\label{mmi}
-I_3(A:B:E) \geq 0,
\end{equation} 
where $I_3(A:B:E) = S_A + S_B + S_E - S_{AB} - S_{AE} - S_{BE} + S_{ABE}$ is the tripartite information. It is quite easy to find counter-examples to  (\ref{mmi}) for generic quantum states, e.g. the 4-party GHZ state, so  (\ref{mmi}) is a special property of holographic states. Rewriting $E_{\text{sq}}$ as a minimization over the tripartite information gives:
\begin{equation}\label{sqen}
E_{\text{sq}}(A:B) = \frac{1}{2}I(A:B) + \frac{1}{2}\min\limits_{\rho_{ABE}}\left[-I_3(A:B:E)\right].
\end{equation}
In general, the state $\rho_{ABE}$ that achieves the minimum in  (\ref{sqen}) need not be geometric in the sense that $E$ may not correspond to a spatial region, and hence MMI need not apply. Fortunately, per our result, the minimum in  (\ref{sqen}) can indeed be achieved by considering only geometric extensions, so the computation is simple: choosing the empty extension $E = \emptyset$ achieves the lower bound $-I_3(A:B:E) = 0$. Note that this corresponds to a purification $\ket{\psi}_{ABEE'}$ where $E$ is empty and $E'$ is any purifier for $\rho_{AB}$. We conclude that the squashed entanglement always saturates its upper bound
\begin{equation}\label{satu}
E_{\text{sq}}(A:B) = \frac{1}{2}I(A:B)
\end{equation}
in holographic systems, giving the holographic dual to the bipartite $E_{\text{sq}}$ as the appropriate sum of minimal surfaces dual to $S_A, S_B,$ and $S_{AB}$. Moreover, taking the squashed entanglement to be a genuine measure of purely quantum correlations, we conclude that correlations within ``nice" holographic, which states are entirely captured by $\frac{1}{2}I$, are dominated at leading order by quantum entanglement.

This technique of using a holographic entropy inequality to make correlation measures computable is quite general: as observed in \cite{Umemoto:2019jlz}, the conditional entanglement of mutual information (CEMI) \cite{PhysRevLett.101.140501}:
\begin{align}
E_I(A:B) &=& \frac{1}{2}\min\limits_{\rho_{ABA'B'}}\bigg[I(AA':BB') - I(A':B')\bigg]
\end{align}
can be evaluated on holographic states by applying MMI: 
\begin{align}
    I(AA':BB') &\geq I(AA':B) + I(AA':B') \\
    &\geq  I(A:B) + I(A':B) \nonumber\\
    &+ I(A:B') +  I(A':B')
\end{align}
after which the positivity of the mutual information implies that the optimal extension is the trivial one, so that $E_I(A:B) = \frac{1}{2}I(A:B)$. The CEMI is another candidate entanglement measure that obeys several pleasing axioms such as monotonicity under LOC, faithfulness, asymptotic continuity, and convexity. The implication for holographic states is similar to that of $E_{\text{sq}} = \frac{1}{2}I$: quantum correlations dominate the total correlations of the state.

We can further use  (\ref{satu}) to obtain the holographic dual for other correlation measures. The saturation of $E_{\text{sq}}$ for holographic states implies that \textit{any} bipartite entanglement measure defined as a constrained minimization of $\frac{1}{2}I(A:B|E)$ over extensions $E$ will reduce to $\frac{1}{2}I(A:B)$, so long as the constraint is compatible with the trivial extension. One example is the $c$-squashed entanglement $E_{\text{sq}}^c$, which is a ``classical'' version of $E_{\text{sq}}$, in the sense that the extension $E$ is constrained to be classical \cite{Tucci:2002ab,Nagel:2003ab,RevModPhys.81.865}:
\begin{equation}
	E_{\text{sq}}^c(A:B) = \min\limits_{\sum_ip_i\rho_{AB}^i\otimes\ket{i}\bra{i}_E}\frac{1}{2}I(A:B|E),
\end{equation}
where the minimization is over all states of the form $\rho_{ABE} = \sum_ip_i\rho_{AB}^i\otimes\ket{i}\bra{i}_E$ with $\sum_ip_i\rho_{AB}^i = \rho_{AB}$. Because this is a constrained minimization, we naturally have $E_{\text{sq}}^c \geq E_{\text{sq}}$, and the lower bound is achieveable with the single-element ensemble $\rho_{AB} \otimes \ket{0}\bra{0}_E$. We must therefore have $E_{\text{sq}}^c = E_{\text{sq}}$, and a corresponding holographic dual to $E_{\text{sq}}^c$.

If $E^\alpha$ is subadditive and its regularization exists, it also collapses to $\frac{1}{2}I$. Specifically, we consider the class of measures $E^{\alpha}_{\text{reg}}$ within the set of $E^{\alpha}$ that equal $E_\text{sq}$ which also satisfy:
\begin{equation}
    E^{\alpha}_{\text{reg}}(A:B) = \lim\limits_{n\to\infty}\frac{1}{n}E^{\alpha}(\rho_{AB}^{\otimes n}), \text{ such that } E^{\alpha} \geq E^{\alpha}_{\text{reg}}.
\end{equation}
Then by the additivity of $E_{\text{sq}}(\rho_{AB}) = \frac{1}{n}E_{\text{sq}}(\rho_{AB}^{\otimes n})$ \cite{doi:10.1063/1.1643788}, we find
\begin{equation}
    E_{\text{sq}} = E^{\alpha} \geq E^{\alpha}_{\text{reg}} \geq E_{\text{sq}}
\end{equation}
and hence $E_{\text{sq}} = E^{\alpha}_{\text{reg}}$.

We are also able to use geometric minimization to find the holographic dual to the multipartite squashed entanglement, introduced in \cite{5075874,Avis_2008}. We first define a multipartite mutual information \cite{Lindblad1973,HORODECKI1994145,PhysRevA.72.032317}:
\begin{equation}\label{multmut}
	I(A_1:\ldots :A_n) = \sum_{i=2}^{n}I(A_1\ldots A_{i-1}:A_{i}).
\end{equation}
The multipartite conditional mutual information $I(A_1:\ldots :A_n|E)$ simply replaces the terms on the RHS of  (\ref{multmut}) with their conditional forms. The multipartite $E_{\text{sq}}$ is then defined to be
\begin{equation}
E_{\text{sq}}(A_1:\ldots:A_n) = \frac{1}{2}\min\limits_{\rho_{A_1\ldots A_nE}}I(A_1:\ldots :A_n|E).
\end{equation}
We can again apply MMI to find:
\begin{align}
	E_{\text{sq}}(A_1:\ldots:A_n) &= \frac{1}{2}I(A_1:\ldots:A_{i}) \nonumber \\
	&- \max\limits_{\rho_{UE}}\sum_{i=2}^{n}I_3(A_1\ldots A_{i-1}:A_{i}:E)  \\
	&\geq \frac{1}{2}I(A_1:\ldots:A_{i}) \nonumber \\
	&- \sum_{i=2}^{n}\max\limits_{\rho_{UE}}I_3(A_1\ldots A_{i-1}:A_{i}:E) \\
	&\geq \frac{1}{2}I(A_1:\ldots:A_n),
\end{align}
and since the lower bound is achievable by the trivial extension, we find that the $n$-party $E_{\text{sq}}$ is equal to one-half the $n$-party mutual information. This confirms the conjecture made in \cite{Umemoto:2019jlz}, and, just as in the bipartite case, immediately implies that the multipartite $E_{\text{sq}}^c$ equal $\frac{1}{2}I$. We can similarly find the dual to the multipartite CEMI \cite{PhysRevLett.101.140501}:
\begin{align}
    E_I&(A_1:\ldots:A_n) \nonumber\\
    &= \frac{1}{2}\min\limits_{\rho_{UU'}}\bigg[I(A_1A_1':\ldots:A_nA_n') - I(A_1':\ldots:A_n')\bigg] \\
    &= \frac{1}{2}\min\limits_{\rho_{UU'}}\left[\sum_{i=2}^n\bigg\{I(A_1A_1'\ldots \nonumber A_{i-1}A_{i-1}':A_iA_i')\right. \\
    &- \left.I(A_1'\ldots A_{i-1}':A_i')\bigg\}\vphantom{\sum_{i=2}^n}\right] \\
    &\geq \sum_{i=2}^nE_I(A_1\ldots A_{i-1}:A_i) \\
    &\geq \frac{1}{2}I(A_1:\ldots:A_n),
\end{align}
and the lower bound is again achievable by the trivial extension, and hence the multipartite $E_I$ is dual to the multipartite $\frac{1}{2}I$.

We stress here that  (\ref{satu}) and other similar equalities are leading order statements in the $1/N$ expansion -- we do not expect equality to hold in the presence of perturbative corrections. Indeed, if this were not the case, then  (\ref{satu}) would imply that holographic states have no classical correlations. Rather, the more accurate statement is that quantum correlations dominate the mutual information, while classical ones are subleading.

\subsection{Axiomatic entanglement measures and the mutual information}
We comment here on the apparent degeneracy of holographic duals to $\frac{1}{2}I(A:B)$. If we work with the heuristic that quantum correlations are the dominant type of correlation in holographic states, then it is perhaps not too surprising that any genuine measure of quantum correlations should saturate at the mutual information at leading order. Therefore, we might, in fact, expect that axiomatic entanglement measures could all equal the mutual information.

One way to formalize this statement is as follows. Consider an entanglement measure $E$ that satisfies the following axioms:
\begin{enumerate}[label=(\alph*)]
\item Normalization: $E(\rho_{AB}) = \log d$ if $\rho_{AB}$ is a maximally entangled state of rank $d$.
\item Monotonicity: for any process $\Lambda$ consisting of local operations and classical communication (LOCC), we have $E(\rho_{AB}) \geq E(\Lambda(\rho_{AB}))$ 
\item Continuity: given two states $\rho,\sigma$ on $\mathcal{H}$, as $||\rho_{AB} - \sigma_{AB}||_1 \to 0$, we also have $\frac{E(\rho_{AB}) - E(\sigma_{AB})}{1+\log \dim \mathcal{H}} \to 0$
\item Regularization: The regularization $E^{\infty}(\rho_{AB}) = \frac{E(\rho_{AB}^{\otimes n})}{n}$ exists for $n \to \infty$.
\end{enumerate}
Any measure that satisfies these axioms also satisfies the bounds \cite{doi:10.1063/1.1495917}:
\begin{equation}\label{sw}
    E_D(\rho_{AB}) \leq E^{\text{reg}}(\rho_{AB}) \leq E_C(\rho_{AB})
\end{equation}
where $E_D$ is the distillable entanglement \cite{PhysRevA.54.3824,PhysRevA.60.173}:
\begin{equation}
    E_D(\rho_{AB}) = \sup\limits_r\left\{r\left|\lim\limits_{n\to\infty}\left[\inf\limits_{\Lambda\in \text{LOCC}}||\Lambda(\rho_{AB}^{\otimes n}) - \Phi_{2^{rn}}||_1 \right]= 0\right.\right\},
\end{equation}
and $E_C$ is the entanglement cost \cite{PhysRevA.54.3824,Hayden_2001}:
\begin{equation}
    E_C(\rho_{AB}) = \inf\limits_r\left\{r\left|\lim\limits_{n\to\infty}\left[\inf\limits_{\Lambda\in \text{LOCC}}||\rho_{AB}^{\otimes n} - \Lambda(\Phi_{2^{rn}})||_1 \right]= 0\right.\right\}.
\end{equation}
Operationally, $E_D$ corresponds to the maximal rate that one can obtain Bell pairs from asymptotically many copies of $\rho_{AB}$, while $E_C$ corresponds to the minimal rate that one can produce asymptotically many copies of $\rho_{AB}$ from Bell pairs. Note that if $E$ is additive, then $E^{\infty}(\rho_{AB}) = E(\rho_{AB})$ and  (\ref{sw}) becomes a bound on $E(\rho_{AB})$ itself. Examples of entanglement measures that satisfy these axioms include the entanglement of formation $E_F$:
\begin{equation}
    E_F(\rho_{AB}) = \min\limits_{\sum_ip_i\ket{\psi_i}\bra{\psi_i} = \rho_{AB}}\sum_ip_iS(\ket{\psi_i}\bra{\psi_i}),
\end{equation}
where the minimization is taken over all pure state decompositions $\{p_i,\ket{\psi_i}\}$ of $\rho_{AB}$, and the relative entropy of entanglement $E_{\text{RE}}$:
\begin{equation}
    E_{RE}(\rho_{AB}) = \inf\limits_{\sigma_{AB}\in\text{Sep}(\mathcal{H}_{AB})}D(\rho_{AB}||\sigma_{AB}),
\end{equation}
where $D(\rho||\sigma) = \tr[\rho\log\rho - \rho\log\sigma]$ is the relative entropy and $\text{Sep}(\mathcal{H}_{AB})$ is the set of separable states on $\mathcal{H}_{AB}$. 

Importantly, $E_{\text{sq}}$ satisfies all 4 axioms and is additive, so:
\begin{equation}
    E_C \leq E_{\text{sq}} \leq E_D.
\end{equation}
If $E_C = E_D$, then we naturally have that all entanglement measures bounded by $E_C$ and $E_D$ are equal. For pure states, $E_C = E_D = S_A$. If $E_C = E_D$ on holographic states (at leading order), then because we have already computed $E_{\text{sq}}$, we would conclude
\begin{equation}
    E_C = E_D = E^{\infty} = E_{\text{sq}} = \frac{1}{2}I(A:B),
\end{equation}
leading to a large degeneracy of holographic duals to $\frac{1}{2}I(A:B)$. Note again that we do not require $E_C = E_D$ exactly, but only approximately. Due to the operational nature of $E_C$ and $E_D$, they are extremely difficult to compute for all the simplest of states and can also exhibit counter-intuitive behaviors, e.g. there exist states which have non-zero entanglement but zero distillable entanglement \cite{Horodecki:1998kf}. Operationally-defined quantities have not received much attention in the context of holography, and a very interesting direction for future study would be a deeper exploration of these objects, leveraging special properties of holographic states to simplify their analysis.

Some recent progress studying $E_D$ in holography is related to the equality of the smooth max and min-entropies for holographic states \cite{Czech:2014tva,Hayden:unpub}. The max and min-entropies of a state $\rho_{AB}$ are defined as:
\begin{align}
    S_{\text{max}} &= \log(\text{rank}(\rho_{AB})), \\
    S_{\text{min}} &= \log(\lambda_{\min}^{-1}(\rho_{AB})), 
\end{align}
where $\lambda_{\min}$ is the minimal eigenvalue of $\rho_{AB}$. The smooth versions of these quantities are defined to be:
\begin{align}
    S_{\text{max}}^\epsilon &= \min\limits_{||\rho-\sigma||_1<\epsilon}\log(\text{rank}(\sigma_{AB})), \\
    S_{\text{min}}^\epsilon &= \max\limits_{||\rho-\sigma||_1<\epsilon}\log(\lambda_{\min}^{-1}(\sigma_{AB})), 
\end{align}
that is, they are the minimal max and maximal min-entropies of states that are in an $\epsilon$-ball around $\rho_{AB}$, where the trace distance is the metric on the space of density matrices. It can be shown \cite{Hayden:unpub,Bao:2018pvs} that evaluating these quantities on holographic states yields:
\begin{align}
    S_{\text{max}}^\epsilon &= S_A + O(\sqrt{S}) \\
    S_{\text{min}}^\epsilon &= S_A - O(\sqrt{S}).
\end{align}
In other words, the smooth max and min-entropies are equal, at leading order, to the von Neumann entropy. This result was then used in \cite{Bao:2018pvs} to show that the one-shot entanglement distillation of a holographic pure state well-approximates $E_D$. In other words, the entanglement spectrum of a holographic state is due almost entirely to maximal entanglement across the entangling surface. This result matches our intuition that quantum correlations are maximal for holographic states, in the sense that $E_{\text{sq}}$ saturates its upper bound. Then at the level of entropies, we can imagine replacing $\rho_{AB}$ with an appropriate number of Bell pairs across the partition, which is quite evocative of $E_C = E_D$.

\section{Discussion}
By using special properties of holographic states, we have shown that optimized correlation measures in holography can be computed, to leading order, by only considering geometric extensions of a given holographic state, provided the states satisfy a set of given assumptions. The most pressing future direction will be an analysis of the assumptions and the degree to which they are reasonable.

We applied our result to confirm the holographic duals for various entanglement measures, including $E_Q$, $E_R$, $E_{\text{sq}}$, $E_{\text{sq}}^c$, and $E_I$. Because all our results are leading order statements, it would be interesting to analyze the form of the quantum corrections that spoil the degeneracy of duals to $\frac{1}{2}I$ and $E_W$.

In our argument, we did not make use of any properties of the optimal purification $\ket{\phi_i}$, besides the fact that it is geometric. Clarifying the properties of this class of purifications would be an interesting future direction. One immediate example is understanding the purifications involved in the $E_P = E_W$ and $E_R = E_W$ proposals. Moreover, we know of several other entanglement measures related to $E_W$ \cite{Dutta:2019gen,Jeong:2019xdr,Bao:2019zqc,Chu:2019etd, Kudler-Flam:2019oru,Tamaoka:2018ned,Kusuki:2019rbk,Kusuki:2019evw,Kudler-Flam:2018qjo,Kudler-Flam:2019wtv,Kusuki:2019zsp}, for which our result will likely be useful in analyzing their connections to $E_P$ and $E_R$.

Although our result confirms several conjectured holographic duals, it alone does not provide candidate duals for the optimized correlation measures. We did not need to use any properties of the bulk spacetime, other than its existence and satisfying certain reasonable assumptions, so generally more tools are necessary. In determining the holographic dual to $E_{\text{sq}}$, we used the MMI property of holographic states. However, MMI is not the only such inequality, and forms one part of the more general ``holographic entropy cone'' \cite{Bao:2015bfa,Hubeny:2018ijt,Cuenca:2019uzx} of inequalities that all holographic states must satisfy, but are not satisfied by generic quantum states. It is likely that these inequalities will similarly prove useful for computing entanglement measures and their holographic duals of higher-partite holographic states. Indeed, one could even \textit{define} entanglement measures of the form  (\ref{opt}) such that the minimization over geometric extensions is simple to compute using holographic entropy inequalities.

Another direction for finding holographic duals is the bit threads formulation of holography \cite{Freedman:2016zud}, which has proven very useful for analyzing $E_P = E_W$ \cite{Bao:2019wcf, Du:2019emy,Harper:2019lff}. We expect that our result, which implies a bit thread configuration exists for the optimal purification in $E^{\alpha}$, will be a useful tool for finding or proving other holographic dual proposals.

A promising direction is to use the saturation of the bipartite squashed entanglement to gain insight into the fine-grained structure of holographic states. In certain cases, the saturation of entropy inequalities or the bounds of information quantities enforces some structure on a state. For example, it is known that the conditional mutual information vanishes on a bipartite state if and only if the state is a so-called ``quantum Markov chain" \cite{Hayden:2004ssa}. We expect that states with maximal squashed entanglement should should have constraints on their form, which in turn imply constraints on the form of holographic states. An example of such a state can be found in \cite{1499048}. Determining such constraints for generic quantum states is likely a difficult task, but it may be possible to, again, leverage properties exclusive to holographic states to make progress in this direction. We leave such explorations to future work.

Finally, it would be extremely interesting to find evidence for or against the conjecture that $E_D = E_C$ for holographic states. Not only would confirmation likely lead to strong constraints on their fine-grained structure, it would also imply a whole host of correlation measures are dual to the mutual information. This also motivates work towards determining the holographic duals of measures such as $E_F$ and $E_{RE}$, whose values could provide evidence in one direction or the other. We leave such explorations to future work.

\begin{acknowledgments}
We thank Ning Bao, Ven Chandrasekar, Masamichi Miyaji, Prathik Rath, Grant Remmen, and Koji Umemoto for useful conversations and helpful comments. We are particularly grateful to Koji Umemoto for a detailed reading and discussions on a draft of this paper. We thank the anonymous referees for suggestions on where our discussion could be improved. We also thank the Yukawa Institute for Theoretical Physics at Kyoto University for hospitality during the workshop YITP-T-19-03 "Quantum Information and String Theory 2019," where several discussions were held that motivated this work.

\end{acknowledgments}

\bibliography{bibliography}

\end{document}